\documentclass[fleqn,twoside]{article}
\usepackage{espcrc2}
\usepackage{graphicx}
\usepackage[figuresright]{rotating}
\newcommand \be {\begin{equation}}
\newcommand \bea {\begin{eqnarray}}
\newcommand \ee {\end{equation}}
\newcommand \eea {\end{eqnarray}}

\newcommand{\bit}{\begin{itemize}}
\newcommand{\eit}{\end{itemize}}

\newcommand{\AmS}{{\protect\the\textfont2
  A\kern-.1667em\lower.5ex\hbox{M}\kern-.125emS}}
\title{ Violation of Lorentz invariance  and dynamical effects in high energy gamma rays}

\author{ P. Castorina\address[dip]{Department of Physics, 
                                                       University of Catania and INFN, Sezione di Catania\\
                                                       Via S. Sofia 64, I 95123, Catania, Italy},
               A. Iorio\address[mit]{Center for Theoretical Physics, 
                                               Massachusetts Institute of Technology\\
                                               77, Massachusetts Avenue, Cambridge MA 02139-4307 U.S.A.},
               D. Zappal\`a\addressmark[dip]}

\begin{document}

\begin{abstract}
The relation between the violation of Lorentz invariance and the dynamical effects in high energy gamma
rays production is discussed.
By using the framework of noncommutative classical electrodynamics, it is shown that full
dynamical calculations are required to put bounds on the Lorentz violating scale by the phenomenological
analysis of these processes as, for example, the synchrotron radiation from the CRAB nebula.
It is observed that an improvement of the present bound on the scale of noncommutativity 
can be obtained only by astrophysical observations of gamma ray spectra in strong magnetic fields
such as pulsars.
\vspace{1pc}
\end{abstract}

\maketitle

\section{INTRODUCTION}


Quantum Gravity (QG) is a work in progress \cite{smolin}. Nevertheless the analysis of  possible
phenomenological effects at low energy with respect to its natural scale, $M_{QG}$, is an active field.
Two of the aspects currently under the most intense investigation, the effects of noncommuting space-time coordinates \cite{nc} and the 
violation of Lorentz invariance \cite{li} \cite{stetc}, have clearly deep connections, although in the literature this is not
always emphasized.

The simplest way to express noncommutativity of space-time coordinates,
\begin{equation}  \label{2}
[x^\mu, x^\nu] = i \theta^{\mu \nu} \;,
\end{equation}
can actually be related to  QG  \cite{doplicher}, while
the breaking of Lorentz invariance is often presented by introducing  
modified dispersion relations, e.g.
\begin{equation}  \label{1}
E^2= {\vec p}~^2 + m^2 +  p^3/M_{QG} \;,
\end{equation}
again motivated by QG or effective field theories \cite{li,coll}.

The role of possible Lorentz violating (LV) effects in ultra high  energy cosmic rays
was used to put bounds on the QG scale $M_{QG}$ by combining the former dispersion relations
together with the energy momentum conservation. 
In this simple and well defined ``kinematic scheme'',  it is easy  to modify the
standard thresholds for decay processes and particle production in collisions
to avoid, for example, the GZK cutoff and to  put limits on $M_{QG}$
from the experimental constraints on the not observed electrodynamic processes, 
as vacuum Cherenkov radiation or $\gamma \rightarrow e^+e^-$ (forbidden in the standard 
case but allowed by the violation of Lorentz invariance \cite{stetc}).

\section{DYNAMICAL AND KINEMATIC ANALYSIS}

For a more comprehensive phenomenological 
analysis, the previous kinematic scheme is not 
enough and one needs a full dynamical calculation of some processes involving 
emission or absorption of radiation.
To this purpose one can consider, for example, the effective field theory introduced in   \cite{coll},  
where the Lorentz violating operators of canonical dimension $\leq 5$  have been 
introduced in the Lagrangian. This produces, on the one hand 
deformed dispersion relations (such as the one in 
Eq. (\ref{1})), on the other hand, a modification of the standard electrodynamics.
However, even though some dynamical assumptions were introduced  in order to put bounds on $M_{QG}$ 
from  high energy astrophysical gamma rays  processes, a complete calculation of such kind of
phenomena in this framework is still missing.
This approach, which gives tight bounds on  $M_{QG}$ \cite{lib}, leaves open the question on  the  
consistency of the dynamical assumptions with  the deformed dispersion relations.

To clarify this point, let us consider the investigated limit on the QG scale obtained 
by the analysis of the  synchrotron radiation from  the CRAB nebula \cite{lib}.
In this case to constrain $M_{QG}$ the following {\it modified} dispersion relations for photons
\begin{equation}
\label{dr1}
E_\gamma^2= {\vec p}~^2 + \xi p^3/M
\end{equation}
and electrons
\begin{equation}\label{dr2}
E^2= {\vec p}~^2 + m^2 + \eta p^3/M\;, 
\end{equation}
are used (where  $1/M_{QG}$ in Eq. (\ref{1}) is replaced  by  $\xi /M$ or $\eta /M$ depending on the particle species,
and $M=10^{19}~GeV$), while the validity of the {\it standard} synchrotron radiation formulas (as the one for the angle of the 
emitted radiation and the critical frequency) are still assumed to hold.

The use of the modified dispersion relations 
within the un-modified dynamics  has been already criticized in \cite{came} and
supported by heuristic arguments in \cite{lib1} (see also the more recent analysis in \cite{lib2}).
However there is no explicit calculation of the synchrotron radiation in the effective field theory
which gives the relations in Eqs. (\ref{dr1}) and (\ref{dr2}),
and, on the other hand, 
it is possible to give other  heuristic arguments that show how the deformed dispersion 
relation for photons in Eq. (\ref{dr1}) produces strong modifications in the synchrotron radiation formulas.

The simplest one is discussed below.
Let us assume, in a classical framework, the following dispersion relation for photons
\begin{equation}  \label{3}
E_\gamma^2= {\vec p}~^2 + \alpha p^n \;,
\end{equation}
with $\alpha >0$ and $n \geq 2$, related to a new wave equation in vacuum. Then, the translation-invariant retarded
Green function can be evaluated \cite{jackson}
\begin{equation}  \label{4}
G(x-x')=\int d^3 p ~d\omega \frac{e^{-ip_\mu(x-x')^\mu}}{\omega^2-p^2(1+\alpha p^{n-2})}
\nonumber
\end{equation}
and the electromagnetic potential generated by a source $J_\mu$ is given by
\begin{equation}  \label{5}
A_\mu(x)=\int d^4x \, G(x-x') J_\mu(x')\; .
\nonumber
\end{equation}
Due to the shift of the poles in Eq. (\ref{4}), the standard retarded Green function \cite{jackson} has the following corrections
\begin{eqnarray} 
G_{\rm corr}(x-x') \sim \alpha\; (t-t')^{n-1} \times && \nonumber\\
\delta^{n-1} \left (t-t' - \frac{|\vec x - \vec x~ ' |}{c} \right )\; , \;\;\;\;\;\;&&
\label{6}
\end{eqnarray}
where $\delta^n$ is the $n-$derivative of the $\delta-$function. In turn, the derivatives of the  $\delta-$function introduce
corrections to the electric and magnetic fields which depend on the derivatives of the acceleration $d^n  \dot \beta/dt^n$.
In the standard case the electric field is proportional to the source acceleration, $E\sim \dot \beta$,
for $n=2$ the correction is proportional to $\ddot\beta$ and so on, 
and the final result is a strong modification 
of the synchrotron radiation spectrum due to the relation in Eq. (\ref{3}).

\section{NONCOMMUTATIVE ELECTRODYNAMICS}

The previous  heuristic argument suggests that the correlation between LV 
terms in the photon dispersion relation and dynamical effects should be treated in a 
well defined framework which  takes consistently  into account both these crucial ingredients.

An example of such a dynamical scheme is the noncommutative electrodynamics 
(NCED) \cite{jack} where the violation of Lorentz invariance and the dynamical 
corrections to the standard  processes are controlled by the same parameters.
The introduction of noncommuting space-time coordinates implies a deformed product between noncommutative fields, 
called Moyal $*$-product \cite{nc}. The Seiberg-Witten map \cite{sw} allows to write the action of NCED in terms
of  the standard product of usual commutative field. At first order in $\theta$  and in the vacuum, one has
\begin{eqnarray}  
\hat{I} = - \frac{1}{4} \int d^4 x  [F^{\mu \nu} F_{\mu \nu}
-\frac{1}{2} \theta^{\alpha \beta} F_{\alpha \beta} F^{\mu \nu}
F_{\mu \nu} + && \nonumber\\
2 \theta^{\alpha \beta} F_{\alpha \mu} F_{\beta \nu}
F^{\mu \nu}]\; , \qquad\qquad\qquad\qquad\qquad\quad&&  
\label{othetamaxwell}
\end{eqnarray}
where $ {F}_{\mu \nu} = \partial_\mu {A}_\nu - \partial_\nu {A}_\mu $.
In the presence of a background magnetic field $\vec b$ the plane wave solutions exist.Waves propagating 
transversely to $\vec b$ enjoy a modified dispersion relation
\begin{equation}\label{8}
\omega/c = k (1 - \vec{\theta}_T \cdot 
\vec{b}_T) \;,
 \nonumber
\end{equation}
while waves propagate  along  the $\vec b$ direction at the standard speed of light $c$. One can also introduce an external source
$J_\mu=(\rho c,\vec J)$ and study the modified Maxwell equations, as done in \cite{ciz}.

In the calculation of the synchrotron radiation (with 
the standard setting of a charged particle moving in the plane $(x,y)$ with speed $\vec \beta$,  $\vec b =(0,0,b),
\theta^{0i}=0, \theta^{ik}= \epsilon^{ijk} \theta_k $  and  $ \theta_k =(0,0,\theta)$,   $\lambda= 2 b \theta$), 
due to the shift of the 
poles in the modified dispersion relation Eq. (\ref{8}), the retarded Green function turns out to be
\begin{eqnarray}
G (\vec{R}, \tau) \sim  \frac{1}{R} \delta(\tau - R / c) \qquad\qquad\qquad\qquad&&\nonumber\\
- \lambda \left( \frac{1 - c \tau / R}{R}
\delta(\tau - R / c) + \frac{\tau}{R} \delta'(\tau - R / c) \right) &&
\label{greenphi} 
\end{eqnarray}
where $\tau = t - t'$. The first term is the standard result. The correction to the electric field is due to the second term
in Eq. (\ref{greenphi})
\begin{eqnarray}
\vec E_{\rm corr}\sim \lambda  \left[\frac{1}{c (1 - \vec{n} \cdot \vec{\beta})} \times \right.\qquad\qquad\qquad
&&\nonumber\\\left.
 \frac{d}{d t'} \left( \frac{1}{c (1 - \vec{n} \cdot
\vec{\beta})} \frac{d}{d t'} \frac{\vec{n} c (t - t')}{(1 -
\vec{n} \cdot \vec{\beta}) R} \right)\right]_{\rm ret}&&
\label{beta2pt} 
\end{eqnarray}
and  contains a term proportional to the derivative of the acceleration.
Let us note that in the previous formulas $\lambda$ is the parameter 
which describes both the violation of Lorentz invariance and the modified dynamics.

For the synchrotron radiation  observed far from the source , in the limit $\beta \to 1$, 
and for frequencies in the region $\omega_0<<\omega<<\omega_c=3 \omega_0 \gamma^3$
(where $\omega_0$ is the cyclotron frequency),
the correction to the spectrum $I(\omega)$ at fixed emission angle, is \cite{ciz}
\begin{equation}\label{ratio}
X \equiv \frac{d I (\omega) / d \Omega}{d I(\omega) / d
\Omega|_{\lambda = 0}} \sim 1 + 
{10 \left (\frac{\omega_0}{\omega}\right )^{2/3} \lambda \gamma^4} 
\nonumber
\end{equation}
and it is potentially large since the coefficient of the parameter $\lambda$ is proportional to $\gamma^4$ 
(with $\gamma=1/\sqrt{1-\beta^2}$)  and depends on the frequency.
Moreover there is a  $O(\lambda)$ correction to the emission angle.

In NCED one  can also evaluate 
the modification to the  Cherenkov radiation (for a medium with magnetic 
permeability $\mu=1$, and electric permeability  $\epsilon=\epsilon(\omega)$)
and, also in this case, the energy radiated per unit distance along the path of the charged particle at fixed frequency,
i.e. $d^2E/(dx d\omega)$, which turns out as 
\begin{equation}\label{12}
\frac{d^2E}{dx~d\omega}\sim \frac{\omega}{c^2} \frac{\epsilon (\lambda-1+\beta^2\epsilon)}{1+\epsilon^2}\; ,
\nonumber
\end{equation}
has a quite different form with respect to the standard case \cite{ciz2}.
Moreover one can show that, due to noncommutative effects, the Cherenkov radiation 
in vacuum  ($\epsilon =1$) is possible if
$\lambda > (1 - \beta^2)/\beta^2 $ and, in this case , the emission angle is fixed by
\begin{equation}\label{13}
\cos^2 \theta=\frac{\beta^2(1+\lambda) -1}{\beta^2(\lambda -1 +\beta^2)}.
\nonumber
\end{equation}

From the previous discussion it seems clear to us 
that, as  in  NCED, one is able to put limits on the LV  parameters by 
dynamical processes involving radiation, only by  consistently considering
the modified dispersion relations {\it and} the modified dynamics. 

\section{BOUNDS ON THE NONCOMMUTATIVITY PARAMETER}

From the discussion in the previous Section  it seems natural to 
ask if it is possible to put bounds directly on the noncommutativity parameter
$\theta$ by the modification of processes involving
high energy gamma rays in NCED, but one need to be careful as NCED is affected with serious problems in 
the quantum phase. These problems are related to a peculiar correspondence between the ultraviolet and
infrared perturbative regimes (see e.g. \cite{ruizruiz}, \cite{esperanza}, \cite{amelia}, \cite{seiberg}). It is still
unclear whether this correspondence is an artifact of the
perturbative calculations or a more fundamental (hence more
serious) problem. For instance in \cite{vaidya} it is shown that
there are noncommutative scalar field theories where the connection is actually
absent. These facts evidently mean that the noncommutative
quantum theory is still a ``work in progress'',  and in the above calculations we used
the (more safe) classical approach  which, in the limit $\theta\to 0$, reproduces 
the standard results \cite{jack}.

With this warning one can  go beyond the use of NCED just as an interesting theoretical laboratory, 
and analyze the possibility that the $O(\theta)$ corrections to the 
processes involved in  high energy gamma rays astrophysics,
may improve the present  bound on the noncommutativity parameter $\theta < (10 {\rm TeV})^{-2}$ \cite{car}.

The parameter $\lambda=2 b \theta$ , which controls the dynamical effects and the LV terms, depends also on
the background magnetic field $b$. Since  galactic and extragalactic magnetic fields are weak, 
there is no improvement on the present bound
by the kinematic modification of the thresholds for the (not forbidden) processes: $\gamma \to e^+ e^- $, 
 $\gamma \gamma\to e^+ e^- $, $e^-\to \gamma e^-$ \cite{noie}.

On the other hand, with the present  limit on $\theta$ the correction to the synchrotron spectrum is
\begin{eqnarray} 
X=\frac{d I (\omega) / d \Omega}{d I(\omega) / d 
\Omega|_{\theta = 0}}  
< 1 + \left ( \frac{\omega_0}{\omega} \right )^{2/3} b \times 10^{-21} \times && \nonumber\\
(E_{\rm electron} (MeV)/(MeV))^4 \qquad\qquad\qquad\qquad&&
\label{14} 
\end{eqnarray} 
where $b$ is the magnitude of the magnetic field expressed in Tesla. For a $20$ TeV electron the correction $X_{corr}$ is
\begin{equation}\label{15}
X_{\rm corr}=\left ( \frac{\omega_0}{\omega} \right )^{2/3} b \times 10^{8} \nonumber
\end{equation} 
and the improvement on the present bound requires strong magnetic fields.

\section{CONCLUSIONS}

According to the brief present analysis one can conclude that: 

1) The bounds on the Lorentz violating scale based on  the purely ``kinematic scheme'' 
reviewed in Stecker's talk \cite{stetc}  are robust because they are independent from the 
underlying dynamics.

2) The bounds from the ``cocktail analysis'',  which mix kinematic and dynamical effects, 
are model dependent and rely on dynamical assumptions.
Such bounds require full dynamical calculations in the effective field theories consistently with the
modified dispersion relations.

3) In order to obtain new
limits on the scale of noncommutativity from electrodynamic processes 
one needs to consider the gamma ray spectra in 
strong magnetic fields as for instance in a pulsar. For this type of analysis one has to take
into account   the whole  
noncommutative effects in the standard chain structure of the radiation process \cite{ho}.

\section{ACKNOWLEDGEMENTS}

The authors thank G. Amelino-Camelia for fruitful discussions.

\end{document}